\documentclass[preprint,aps,superscriptaddress,nofootinbib,tightenlines,showpacs]{revtex4}

\usepackage{epsfig}
\usepackage{amssymb}
\usepackage{color}

\def\OMIT#1{{}}

\newcommand{\beq}{\begin{equation}}
\newcommand{\eeq}{\end{equation}}
\newcommand{\beqa}{\begin{eqnarray}}
\newcommand{\eeqa}{\end{eqnarray}}

\newcommand{\nn}{\nonumber}

\begin{document}

\preprint{\vbox{
\hbox{NT@UW-14-14}
\hbox{INT-PUB-14-020}
}}
\bigskip

\title{Two-Particle Elastic Scattering in a Finite Volume Including QED}
\vskip 0.8cm

\author{\bf Silas R.~Beane}
\affiliation{Department of Physics, University of Washington, Seattle, WA 98195}
\author{\bf Martin J.~Savage}
\affiliation{Institute for Nuclear Theory, University of Washington, Seattle, WA 98195}

\vphantom{}
\vskip 1.8cm
\begin{abstract} 
\noindent 
The presence of long-range interactions violates a condition necessary
to relate the energy of two particles in a finite volume to their
S-matrix elements in the manner of L\"uscher.  While in infinite
volume, QED contributions to low-energy charged particle scattering
must be resummed to all orders in perturbation theory (the Coulomb
ladder diagrams), in a finite volume the momentum operator is
gapped, allowing for a perturbative treatment.  The leading QED
corrections to the two-particle finite-volume energy quantization
condition below the inelastic threshold, as well as approximate
formulas for energy eigenvalues, are obtained.  In particular, we
focus on two spinless hadrons in the $A_1^+$ irreducible
representation of the cubic group, and truncate the strong
interactions to the s-wave.  These results are necessary for the
analysis of Lattice QCD+QED calculations of charged-hadron
interactions, and can be straightforwardly generalized to other
representations of the cubic group, to hadrons with spin, and to
include higher partial waves.

\end{abstract}


\maketitle


\vfill\eject

\section{Introduction}
\noindent 
Lattice QCD (LQCD) calculations of the properties of the lowest-lying
mesons are reaching the level of accuracy where it is necessary to
consider the strong interactions in the context of the full Standard
Model.  In particular, hadronic spectra and other hadronic observables
are now being calculated in the presence of both isospin violation
from the light-quark masses and Quantum Electrodynamics
(QED)~\cite{Blum:2007cy,Basak:2008na,Blum:2010ym,Portelli:2010yn,Portelli:2012pn,Aoki:2012st,deDivitiis:2013xla,Borsanyi:2013lga,Drury:2013sfa,Horsley:2013qka,Borsanyi:2014jba}.
QED plays a critical role in the stability and structure of nuclei,
and therefore first principles calculations of nuclear structure
require the inclusion of the electromagnetic (EM) interactions among
quarks.  Due to computational resource limitations, LQCD calculations
of nuclei remain at an early stage, with calculations of the binding
energies of systems with up to five nucleons and hyperons currently
being performed at unphysical light-quark
masses~\cite{Beane:2009py,Yamazaki:2009ua,Beane:2010hg,Inoue:2010es,Inoue:2011pg,Beane:2011iw,Yamazaki:2011nd,Yamazaki:2012hi,Yamazaki:2012fn,Beane:2012vq}.
While the time is not yet ripe for the inclusion of QED in nuclear
calculations, there are two-body scattering processes that can now be
calculated with high accuracy in LQCD and where Coulomb corrections
are relevant, for instance $\pi^+\pi^+$.  Therefore, formalism that
allows for the systematic calculation of electromagnetic corrections
to two-body interactions in a finite volume (FV) is required.

The extraction of hadronic interactions from Lattice QCD calculations is
more complicated than determining the spectrum of stable hadrons.  The
Maiani-Testa theorem~\cite{Maiani:1990ca} demonstrates that S-matrix
elements cannot be directly extracted from infinite-volume
Euclidean-space Green's functions except at kinematic thresholds.
While discouraging from the viewpoint of nuclear physics, where a
central objective is determining the forces between nucleons, hyperons
and other hadrons, it is clear from its statement that the theorem can
be evaded with FV calculations.  The essential formalism that enables
extraction of continuum S-matrix elements describing two-body elastic
scattering from measurements of two-body energies in a finite spatial
volume has been known for decades in the context of non-relativistic
quantum mechanics~\cite{Huang:1957im} and, for two spinless particles,
was extended to quantum field theory by
L\"uscher~\cite{Luscher:1986pf,Luscher:1990ux}.  The energy of two
particles in a FV depends in a calculable way upon their elastic
scattering amplitudes, and their masses, for energies below the
inelastic threshold. A fundamental assumption in this formalism is
that the two particles experience only finite-range interactions, such
that the typical interaction length scale is well-contained within the
spatial volume.  Recently, L\"uscher's formalism has been extended to
coupled-channels systems (i.e.~channels that are coupled in
infinite-volume), and to systems comprised of particles with non-zero
spin~\cite{Detmold:2004qn,He:2005ey,Liu:2005kr,Bernard:2008ax,Lage:2009zv,Bernard:2010fp,Ishizuka:2009bx,Briceno:2012yi,Hansen:2012tf,Guo:2012hv,Li:2012bi,Briceno:2013bda,Briceno:2014oea}.
Further, the FV formalism describing nucleon-nucleon (NN) systems with
arbitrary CM momenta, spin, angular momentum, isospin and twisted boundary
conditions has been developed, providing the quantization conditions
(QCs) for the energy eigenvalues in irreducible representations
(irreps) of the FV symmetry groups~\cite{Briceno:2013lba}.  Efforts to
account for the exponentially-suppressed effects of the finite range
of the interactions have also been
made~\cite{Bedaque:2006yi,Sato:2007ms}.

\begin{figure}[!t]
  \centering
     \includegraphics[scale=0.625]{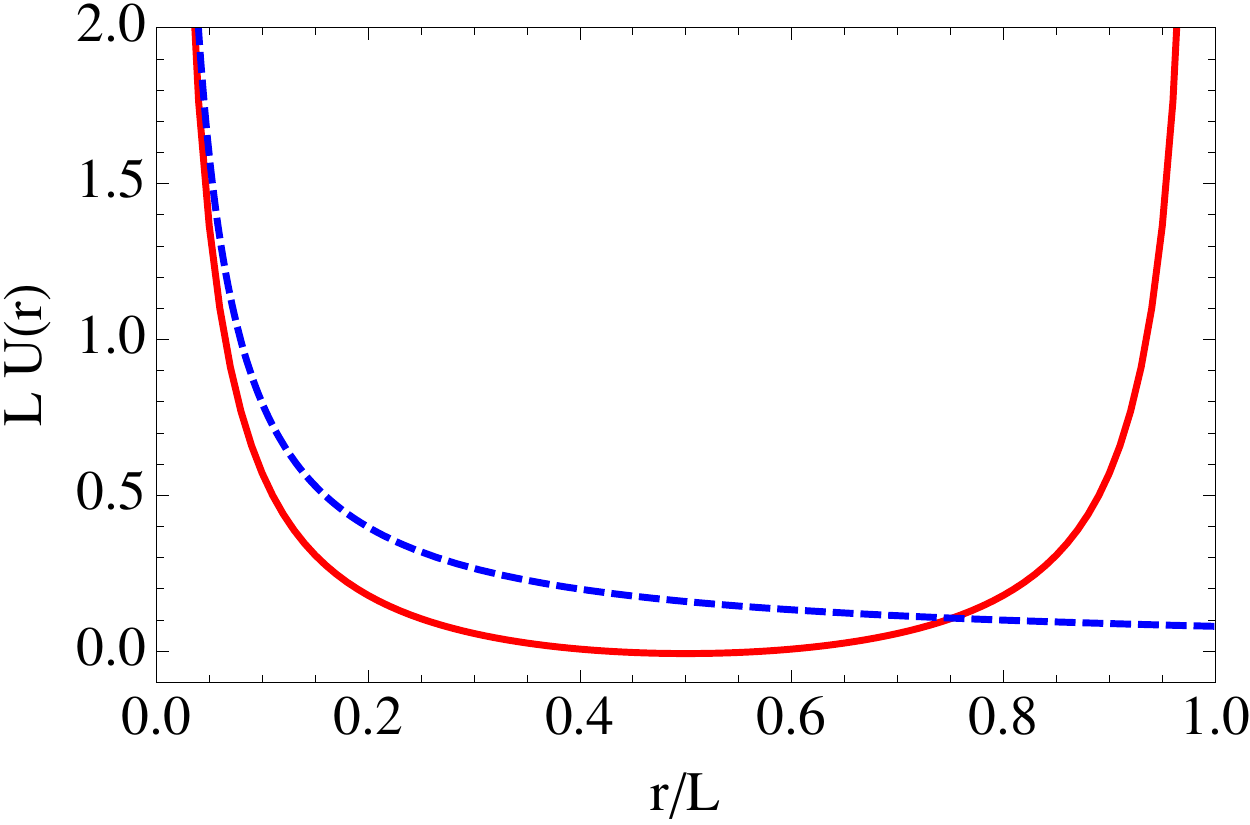} \qquad  \qquad  \includegraphics[scale=0.3]{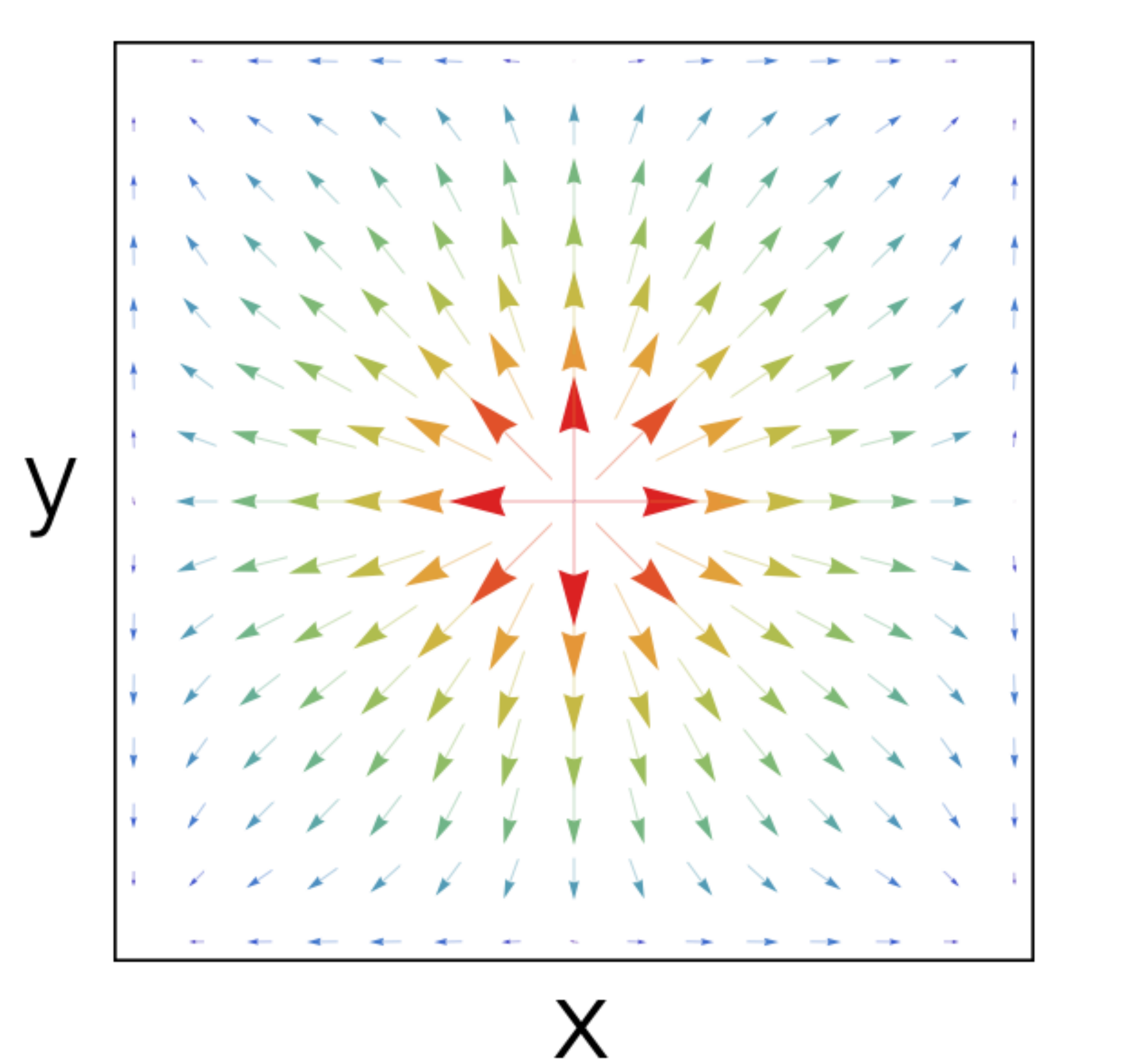}
     \caption{
     The left panel shows 
     the FV Coulomb potential energy between unit charges along an axis of a cubic volume (solid red curve) 
     obtained from Eq.~(\ref{eq:FVcp}), and the infinite-volume Coulomb potential (dashed blue curve)~\protect\cite{Davoudi:2014qua}.
     The right panel shows the FV electric field in the $z=0$ plane due to a point charge located at the center of the cube.
     }
  \label{fig:coulpot}
\end{figure}

At a fundamental level, the inclusion of QED into LQCD calculations
poses a theoretical challenge, as the long-range nature of the
interaction is truncated and modified by the boundary of the volume.
In particular, Ampere's law and Gauss's law cannot be satisfied with a
QED gauge field subject to periodic boundary conditions
(PBCs)~\cite{Hilf1983412,Duncan:1996xy,Hayakawa:2008an,Davoudi:2014qua}.
A uniform background charge density can be introduced to circumvent
this problem, a procedure which is equivalent to removing the zero
modes of the photon.  That is, the Coulomb
potential energy between charges, $e$, in a cubic spatial volume with
the zero modes removed, is
\begin{eqnarray} 
U( {\bf r} ,L) 
& = &  
{\alpha \over \pi L}
\sum_{{\bf n}\ne {\bf 0}}
{1\over |{\bf n}|^2}
e^{i 2\pi {\bf n}\cdot {\bf r}\over L}
\label{eq:FVcp}
\end{eqnarray}
where $\alpha=e^2/4\pi$, ${\bf n}$ are triplets of integers and $L$ is
the spatial extent of the cubic volume.  The FV Coulomb potential can
be seen in comparison with the infinite-volume potential in
Fig.~\ref{fig:coulpot} (left panel). A cross section of the FV
electric field due to a point charge in the center of the volume is
show in Fig.~\ref{fig:coulpot} (right panel).  Given the large density
of momentum states in typical lattice volumes, the removal of the zero
modes will not change the desired infinite-volume values of calculated
observables~\footnote{The FV modifications to the values of
  counterterms in a low-energy effective field theory of QCD will scale as
  $\sim e^{-L/r}$, where $r$ is the typical scale of the strong
  interactions.  }.

In the absence of QED, there is a clear separation of the FV artifacts
into those that behave as power laws in $L$, and those that are
exponentially suppressed in $L$.  The latter are governed by the
longest correlation length in the volume, which, in chiral
perturbation theory ($\chi$PT) and nucleon-nucleon effective field
theory (NNEFT), is the pion Compton wavelength.  In contrast, the QED
FV effects behave as a power law, which means that the energy
eigenvalues of two charged hadrons will be modified in the same way by
their self interactions and by their interactions with each other.
Therefore, unlike the case with only short-range forces, in the
presence of photons, the kinematics of ``scattering processes'' in
lattice calculations also receive power-law modifications in the FV.

The separation of QED effects from strong interaction effects in
scattering processes has a long history.  However, it is convenient to
use effective field theory (EFT) technology, and its associated
power-counting, in deriving the QED corrections to the FV QCs, the
solution of which provides the energy eigenvalues.  Generally, for
low-energy charged-particle scattering processes, the Coulomb
interaction is included nonperturbatively through a resummation of
ladder diagrams.  In an infinite volume this is necessary because the
scale of the Coulomb bound state is set by the ``Bohr'' radius,
$(\alpha M)^{-1}$, and interactions with momenta that probe the
binding energy of the system are nonperturbative in $\alpha$.  In FV,
the non-perturbative treatment would appear to be quite involved due
to the proliferation of increasingly complex integer sums.  However,
in the spatial lattice volume, $L^3$, the momentum operator is gapped,
with a scale that is set by $1/L$, and not by the inverse Bohr radius.
Therefore, there is a range of volumes in which the QED interactions
can be treated in perturbation theory in a loop expansion, leading to
a significant simplification in the corrections to L\"uscher's QCs.
Another energy scale that must be considered is the inelastic
threshold, set by the lowest photon energy in the FV, $E = 2\pi/L$.
Given that there are no zero-modes in the FV, by construction, some of
the infrared (IR) issues that are usually encountered in QED are absent.
As expected, this threshold dictates the kinematical region of validity of
the truncation of the QC to two-body states.

This paper is organized as follows. In Section~\ref{sec:rev}, we
review the basic EFT results that allow for a separation of the QCD
and QED interactions in the elastic scattering of two charged hadrons
in the continuum.  These results form the basis of the FV
generalization.  QED modifications to the FV QCs that provide the
energy eigenvalues of the $A_1^+$ cubic irrep, truncated to s-wave
interactions, are the subject of Section~\ref{sec:EMcorrs}.  First,
the modifications to the scattering process kinematics due to FV
self-energy shifts is considered, then the truncated QC is determined.
In the limit of small scattering lengths compared to $L$, perturbative
expressions for the energy eigenvalues are derived.  Furthermore, the
QED corrections to the energy of a bound state (when one exists) are
determined.  Requisite integer sums are provided in the Appendix.

\section{Coulomb Scattering}
\label{sec:rev}
\noindent 
QED contributions to two-particle interactions in a FV will be
considered in the context of the pionless
EFT~\cite{Kaplan:1998tg,Kaplan:1998we,Kaplan:1998sz,vanKolck:1998bw,Chen:1999tn,Phillips:1999hh,Beane:2000fi,Bedaque:2002mn}.
The effective range expansion (ERE), which describes the low-energy
strong interactions between two hadrons, emerges naturally from the
pionless EFT, and it was shown by Bethe~\cite{Bethe:1949yr} how the
ERE is modified in the presence of Coulomb interactions.  Bethe's
analysis was reformulated in EFT by Kong and
Ravndal~\cite{Kong:1999sf}, and as this formalism plays a central role
in the calculations that follow, it is helpful to review its salient
features.

The T-matrix describing the QED interactions of two spinless charged
particles of mass $M$, charge $e$, carrying equal but opposite
momentum ${\bf p}$, and in the absence of strong interactions, has a
partial-wave expansion of the form
\begin{eqnarray}
T_{C} & = & 
-{4\pi\over M}
\sum_l\ (2l+1) 
\ {e^{i 2 \sigma_l}-1 \over 2 i p}\ 
P_l(\cos\theta)
\ \ \ ,
\end{eqnarray}
where $p=|{\bf p}|$ and $\sigma_l = {\rm arg}\;\Gamma(1+l+i\eta)$.  $l$ is the angular
momentum of the scattering channel, $\eta = \alpha M/(2 p)$, and
$\theta$ is the center-of-mass (CoM) scattering angle.  The strong
interactions between two hadrons below the t-channel cut in an s-wave
can be described by an EFT of four-hadron operators.  The effects of
these operators can be encapsulated, for the purposes of this work, by
a single interaction (a pseudo-potential) with a coefficient $C(E^*)$,
which is an analytic function of the CoM energy $E^*$~\footnote{At the level of the non-relativistic Lagrange density, expressed as a
gradient expansion of local operators built out of a field $\psi$, it
is straightforward to show, using equations of motion and integrating
by parts, that~\cite{Birse:1998dk,Beane:2000fi}
\begin{eqnarray}
-{\hat\theta} 
\ \psi^T (\overleftarrow\nabla - \overrightarrow\nabla)^2 \psi
& = & 
4 M \ {\hat\theta}\ 
\left[ i \partial_0 + {\nabla^2\over 4 M}\right]  \psi^T  \psi
\equiv 
4 M \ {\hat\theta}\ {\cal O}_{E^*} \psi^T  \psi
\ \ \ ,
\end{eqnarray}
where ${\hat\theta}$ is an arbitrary operator, and terms that are total derivatives
are not shown. The operator ${\cal O}_{E^*}$ when acting on the two-particle operator
simply yields the non-relativistic center-of-mass energy, $E^*$. 
}. 

Treating $C(E^*)$ nonperturbatively by summing all bubble diagrams with a
$C(E^*)$ insertion at each vertex, and using dimensional
regularization (DR) to regulate ultraviolet divergences, the T-matrix including
the strong and the leading QED interactions is
\begin{eqnarray}
T_{SC} & = & 
C_{\eta (p)}^2\ { C(E^*) e^{i 2 \sigma_0} \over 1 - C(E^*) J_0^{\infty}(E^*)}\ =\ - {4\pi\over M} {e^{2i\sigma_0}\over p \cot\delta - i p}
\ \ \ ,
\label{eq:TSCere}
\end{eqnarray}
where $\delta$ is the s-wave phase shift.
$J_0^\infty(E^*)$ is the 
${\bf r}={\bf 0}$ to ${\bf r}={\bf 0}$ 
Green's function including QED interactions,  
and can be written as
\begin{eqnarray}
J^\infty_0(E^*) \ =\  M\!\int\!{d^3 q\over (2\pi)^3} {C^2_{\eta(q)}\over p^2 - q^2 + i\epsilon} \ ,
\label{Cr2b}
\end{eqnarray}
and  $C_{\eta(p)}$ is the Coulomb corrected vertex resulting from the resummation of Coulomb ladder diagrams, with a square
given by
\begin{eqnarray}
C^2_{\eta(p)} \ =\ {2\pi\eta(p)\over e^{2\pi\eta(p)} - 1} 
\ \ \ .
\label{Cr3b}
\end{eqnarray}
The parameter $\eta\sim \alpha/v$, where $v$ is the relative velocity of the two hadrons, governs the viability of 
QED perturbation theory
and therefore, as pointed out above, for momenta of order $\alpha M$, $\eta\sim 1$ and Coulomb
ladders must be treated to all orders in $\alpha$ and resummed.

\begin{figure}[!t]
  \centering
     \includegraphics[scale=0.22]{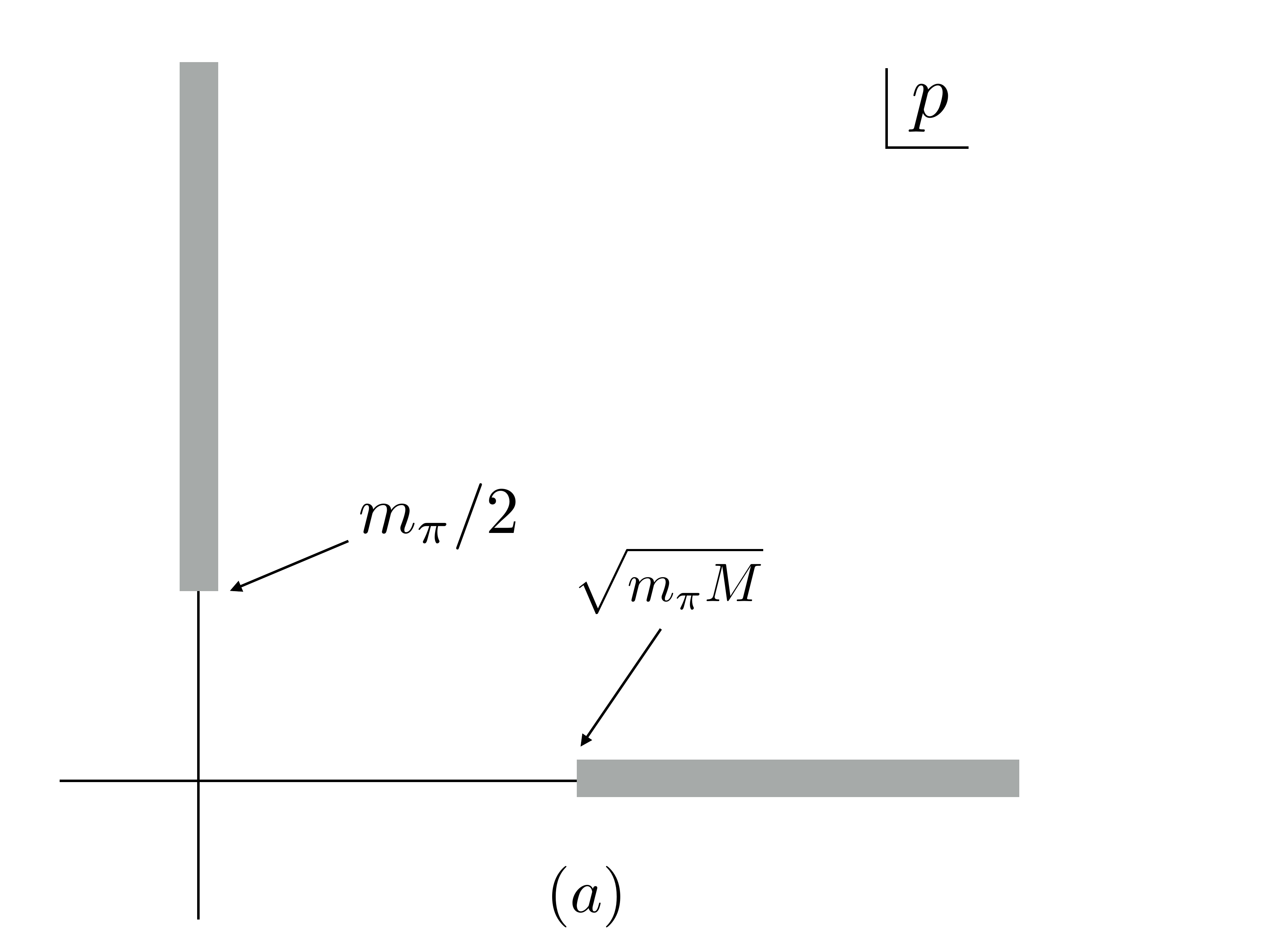}     \includegraphics[scale=0.22]{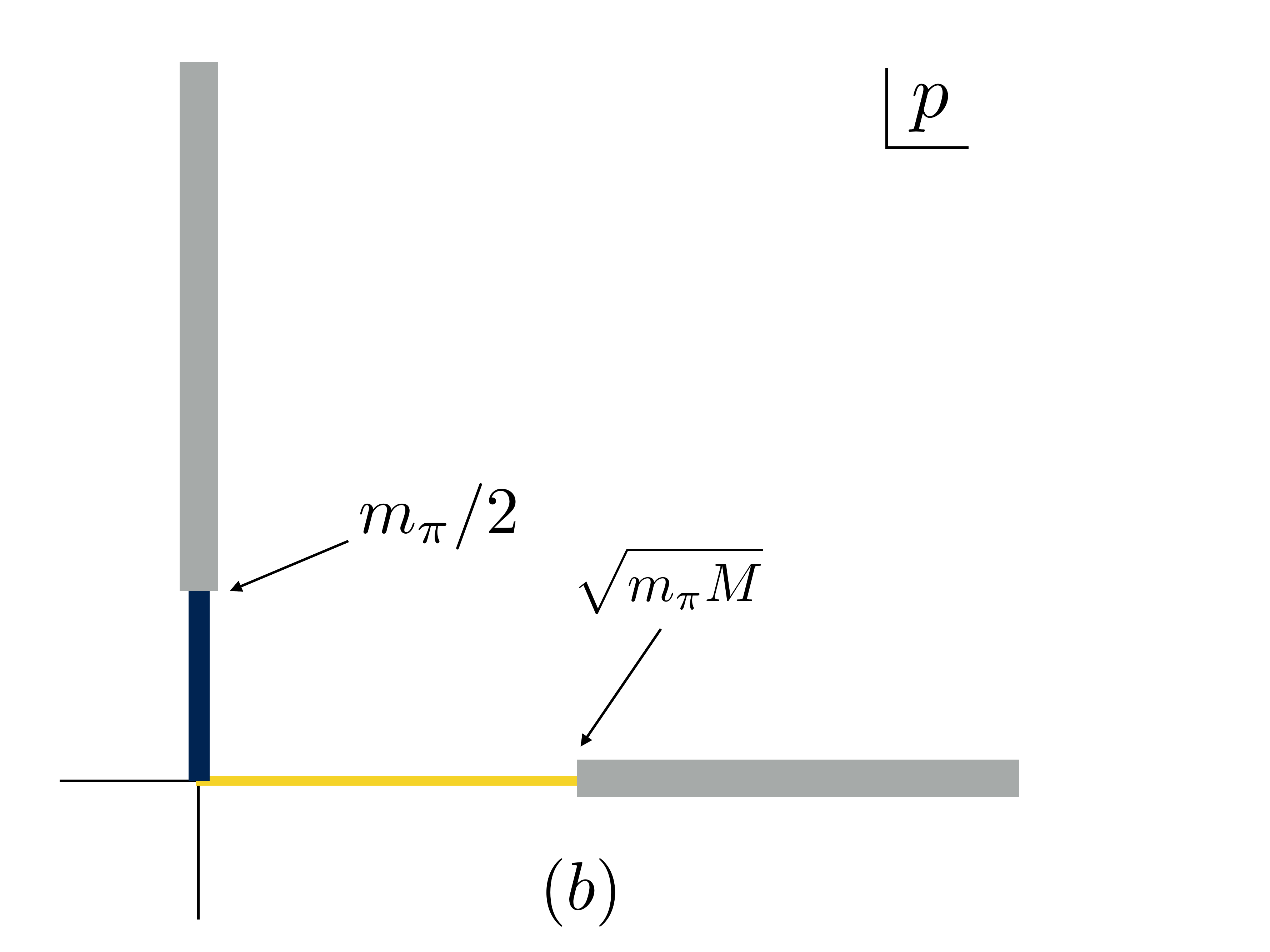}
     \caption{The analytic structure of the scattering amplitude in the complex $p$ plane (a) without QED and (b) with QED. 
The imaginary axis exhibits the QCD t-channel cut with its threshold at $m_\pi/2$, 
while the real axis gives the inelastic pion-production cut with its threshold at $\sqrt{m_\pi M}$. In the presence of
QED, both the t-channel cut (dark blue) and the inelastic cut (yellow) begin at the origin.}
  \label{fig:panalyticstrong}
\end{figure}

Decomposing $J^\infty_0$  into finite and divergent parts, $J_0^{{\it fin}}+J_0^{{\it div}}$, leads to~\cite{Kong:1999sf}
\beq
J_0^{\it fin} \ = \ M\!\int\!{d^3 q\over (2\pi)^3} 
     {C^2_{\eta(q)}\over q^2}{p^2\over p^2 - q^2 + i\epsilon} \ =\ -\frac{\alpha M^2}{4\pi}\; H(\eta ) \ ,
\label{Cr4}
\eeq
where
\beq
H(\eta) \ =\  \psi(i\eta) + {1\over 2i\eta} - \ln(i\eta) \ ,
\label{Cr5}
\eeq
with $\psi$  the logarithmic derivative of the Gamma function.
Using DR with modified minimal subtraction ($\overline{MS}$) 
in $n=4-2\epsilon$ dimensions,~\footnote{
The power counting in the EFT is manifest in the PDS scheme~\cite{Kaplan:1998tg,Kaplan:1998we}. 
Here for simplicity we use $\overline{MS}$, which obscures the strong power counting, but does not change it.
}
the divergent part becomes
\beq
J_0^{div} \ =\  - M\!\int\!{d^3 q\over (2\pi)^3} 
{C_{\eta(q)}\over q^2} \ =\  {\alpha M^2\over 4\pi}\left[{1\over\epsilon} + \ln\left({\mu\sqrt{\pi}\over\alpha M}\right)
                  + 1 - {3\over 2}\gamma_E \right]  \ ,
\label{Cr6}
\eeq
where $\mu$ is the renormalization scale introduced in $n$ dimensions, and
$\gamma_E$ is Euler's constant.
The expression for $T_{SC}$ in Eq.~(\ref{eq:TSCere}) then leads to
\beqa
C_{\eta(p)}^2\; p \cot\delta  + \alpha M h(\eta) 
& =& -\frac{4\pi}{M C(E^*)} \ +\ 
 \alpha M\left[{1\over\epsilon} + \ln\left({\mu\sqrt{\pi}\over\alpha M}\right)
                  + 1 - {3\over 2}\gamma_E\right] 
\label{Cr7}
\eeqa
where 
\beq
{\rm Im} H(\eta) \ =\ \frac{C_{\eta(p)}^2}{2\eta} \ \ \ \ {\rm and} \ \ \ \ 
{\rm Re} H(\eta) \ \equiv \ h(\eta) 
\ ,
\label{Cr8}
\eeq
have been used. As Bethe showed, the left-hand side of Eq.~(\ref{Cr7})  admits an ERE of the form
\beqa
C_{\eta(p)}^2\; p \cot\delta  + \alpha M h(\eta) 
   & =& - {1\over a_C} \ +\ \frac{1}{2}r_0 p^2\ + \ \ldots                 
\label{Cr7b}
\eeqa
where $a_C$ is the Coulomb-corrected scattering length and $r_0$ is
the effective range. The presence of the extra term on the left-hand
side can be understood based on the analytic structure of the scattering
amplitude (see Fig.~\ref{fig:panalyticstrong}). As the t-channel cut begins
at the origin when photons are present, this term removes this cut from
the scattering amplitude, thus leaving an expression that is analytic in $p^2$
(neglecting radiation) and which consequently admits an effective range expansion. 
While the inelastic threshold is at $p=0$, this cut is suppressed by powers
of $\alpha$ compared to the t-channel cut.

Matching the right-hand sides of Eq.~(\ref{Cr7}) and Eq.~(\ref{Cr7b}) is achieved through
renormalization~\cite{Kaplan:1998tg,Kaplan:1998we,Kong:1999sf}.
Rather than use $\overline{MS}$ to subtract the $1/\epsilon$ pole, 
a slightly modified scheme, denoted $\overline{MS}_{FV}$, is used, which corresponds to subtracting
\beq
{\alpha M^2\over 4\pi}\left[{1\over\epsilon} -\frac{\gamma_E}{2} + 1 +\ln\frac{\sqrt{\pi}}{2} \right] \ .
\label{Cr10}
\eeq
In this scheme, which is convenient for the FV calculations to follow, 
the ERE can be described by renormalized coefficients,
\beq
-{4\pi\over MC(p;\mu)} \ +\  \alpha M \left[\ln\left({2\mu\over\alpha M}\right) - \gamma_E \right]  \ =\ {1\over a_C} \ +\ \frac{1}{2}r_0 p^2\ + \ \ldots   \ ,
\label{Cr11}
\eeq
where $C(p;\mu)=C_0(\mu )+C_2(\mu)p^2+\ldots$ is the renormalized strong interaction.

The analysis of this section is appropriate for the interactions of like-charged hadrons, such as proton-proton scattering.
In the case of hadrons with opposite charges, the kinematic factor $\eta$ changes sign, $\eta = -\alpha M/(2 p)$, and 
$H(\eta)$ becomes
\beq
\overline{H}(\eta) \ =\  \psi(i\eta) + {1\over 2i\eta} - \ln(-i\eta) \ ,
\label{Cr5Hopposite}
\eeq
%

\section{Finite Volume Coulomb Scattering}
\label{sec:EMcorrs} 

\subsection{Power Counting and Kinematics}
\noindent 
In a cubic spatial volume with PBCs, a free particle can carry
momentum ${\bf p}={2\pi {\bf n}/L}$, where ${\bf n}$ is a triplet of
integers.  In the absence of zero modes, the momentum carried by a
photon is restricted to $k \ge {2\pi/L}$ and the relevant size of
$\eta$ in the FV is $\eta\sim \alpha M L$, which implies that for $M L
\ll 1/\alpha$, QED interactions can be treated perturbatively in
$\alpha$.  Of course, $\eta$ grows with the spatial volume and, for a
given $M$, there is a critical value of $L$ at which
perturbation theory breaks down and the Coulomb ladders must be
resummed to all orders, as in the continuum.  In addition, LQCD
calculations have volumes large enough so that $M \gg 1/L$, and this
limit will also be assumed throughout this analysis. Note that due to the
absence of the zero mode, the inelastic threshold of the two-hadron state,
which is set by the two hadrons recoiling against a photon, is at 
$\sqrt{2\pi M/L}+{\cal O}(1/M)$.

The power-law nature of the expansion parameter leads to various
subtleties.  In the absence of QED, hadron self energies contain FV
corrections that are exponentially suppressed by the dimensionless
parameter $m_\pi L$, and therefore, neglecting these
corrections, the kinematics in the FV are the same as in the
continuum.  This is no longer the case in the presence of QED as the
hadron masses have power-law volume 
dependencies~\cite{Hilf1983412,Duncan:1996xy,Hayakawa:2008an,Davoudi:2014qua}.

The total CoM energy of the two-hadron system can be written as 
 $E^* =  2 M^L + T^{* L}$,  where $T^{* L}$ is the CoM kinetic energy, and $M^L$ is the mass of the single hadron, in the FV.
The ERE, while usually written in terms of an expansion in square of the hadron three momentum,
is an analytic function of $E^*$ below the inelastic threshold,
and with the FV shift in the hadron mass(es), is evaluated at a shifted value of the kinetic energy in the FV,
\begin{eqnarray}
p\cot\delta & = & -{1\over a_C} + {1\over 2} r_0 p^2 + r_1 p^4 \ +\ ...
\ =\ -{1\over a_C} + {1\over 2} r_0 M T^* + r_1 M^2 T^{* 2}\ +\ ...
\nonumber\\
& = & 
-{1\over a_C} + {1\over 2} r_0 M (E^* - 2 M) + r_1 M^2 (E^* - 2 M)^2 \ +\ ...
\nonumber\\
& \rightarrow & 
-{1\over a_C^\prime} + {1\over 2} r_0^\prime M T^{* L} + r_1^\prime M^2 (T^{* L })^2 \ +\ ...
\ \ \ \  ,
\end{eqnarray}
where $r_1$ is the shape parameter.
The primed scattering parameters, that are required to describe the FV two-point function,
are defined by 
\begin{eqnarray}
{1\over a^\prime_C} & = & {1\over a_C} 
-   {\alpha\  r_0 M\,{\cal I}\over 2\pi L}
\ +\ {\cal O}(\alpha^2;\alpha/L^2)
 \ \ , \ \ 
r_0^\prime \ =\  r_0  + { 4 \ \alpha\ r_1  M \ {\cal I}\over \pi L}
\ +\ {\cal O}(\alpha^2;\alpha/L^2)
\, ,
\end{eqnarray}
with similar modifications to the terms that are higher order in the
ERE.  In these shifted ERE parameters, the single-particle FV
corrections of Refs.~\cite{Hayakawa:2008an,Davoudi:2014qua} have been
used, and ${\cal I}\sim-8.913632$ is an integer sum detailed in the
Appendix.

Up to this point, the discussion has been focused on the dynamics of
point-like particles, but as this work is relevant to LQCD
calculations, the effect of compositeness must be considered.  In
Ref.~\cite{Davoudi:2014qua}, the EFT describing the low-energy
dynamics of hadrons was used to determine the FV corrections to hadron
masses in LQCD calculations, in which the effect of compositeness,
manifesting itself through a hierarchy of electromagnetic multipole
interactions and other multi-photon gauge-invariant interactions, was
made explicit.  These one-body QED interactions, beyond the electric
charge, will also contribute to energy eigenvalues of two hadrons,
electrically charged or neutral.  For spinless hadrons, the leading
interaction beyond the charge is from its charge radius.  Given that
the charge radius is proportional to the square of the momentum
carried by the photon, the leading effect of the charge radius is to
provide an constant additive renormalization of $C(E^*)$, which is the
same in finite and infinite volume.  Further, this contribution cannot
be isolated from the experimental scattering data without a model
dependent subtraction, or with a explicit calculation of
the low-momentum transfer contribution using EFT.  Therefore, in what
follows, the leading contribution from the structure of spinless
hadrons is already included in the definition of the scattering
parameters, and the comparison with experiment should not remove this
contribution from the experimental data prior to comparing.  The
analysis that follows does not make explicit the contribution from the
hadron charge radius, but one should keep in mind that it is implicit.

\subsection{Quantization Condition including QED}
\begin{figure}[!t]
  \centering
     \includegraphics[scale=0.4]{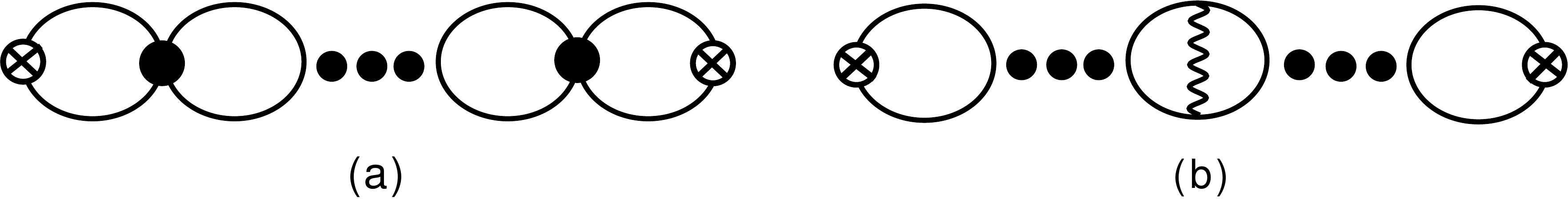}
     \caption{
     Feynman diagrams contributing to the FV two-point function.
     Diagram (a) is one of the bubble diagrams resulting from the strong interactions,
     while diagram (b) is one of the diagrams at ${\cal O}(\alpha)$ from the exchange of a Coulomb photon
     (that becomes one of the Coulomb ladder diagrams in infinite-volume).
     }
  \label{fig:BubblesAndCoulomb}
\end{figure}

\begin{figure}[!t]
  \centering
     \includegraphics[scale=0.31]{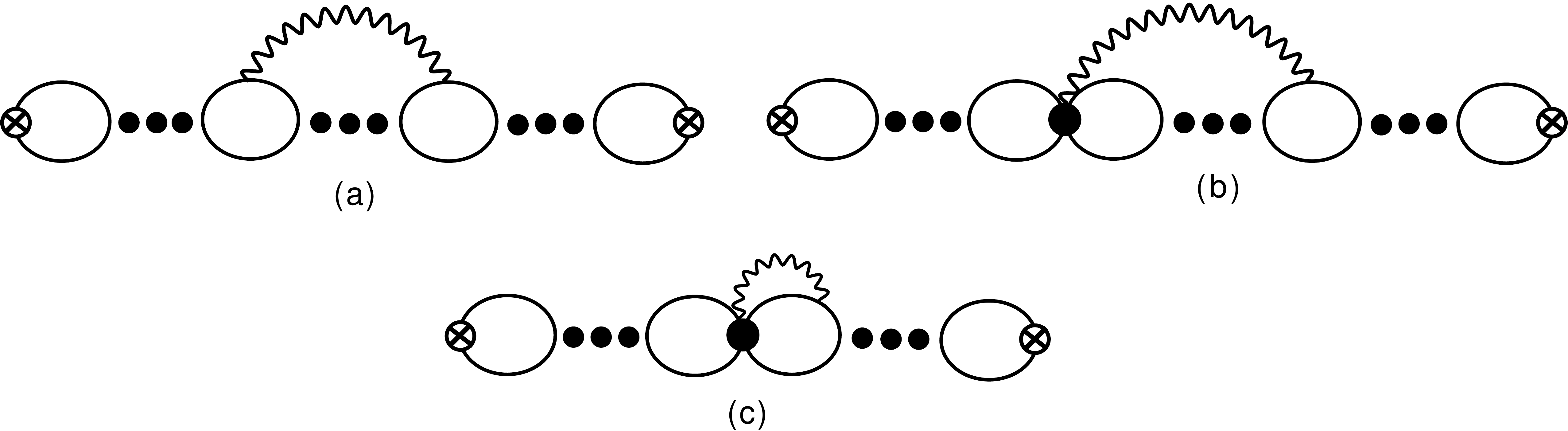} 
     \caption{
     Feynman diagrams contributing to the FV two-point function but which 
     are suppressed in the IR compared to the Coulomb ladder diagrams.
       }
  \label{fig:BubblesAndRadiationA}
\end{figure}

\noindent
The truncated  QC that determines the $A_1^+$ FV energy eigenvalues 
can be  determined by the singularities of the FV two-point function.
In general, the ${\cal O}(\alpha)$ corrections to the two-point function 
result from the sum of all diagrams with a single insertion of a photon and the related counterterms,
examples of which are shown in Fig.~\ref{fig:BubblesAndCoulomb} and Fig.~\ref{fig:BubblesAndRadiationA}.
Consider the correlation function between a source, $S^\dagger $ and a
sink, $S$, where $S^\dagger, S $ couple to two hadrons in an s-wave.
Denoting the contribution to this two-point function from the sums of
bubbles shown in Fig.~\ref{fig:BubblesAndCoulomb} as $J_0^L(E^*)$, and
the (generally) energy-dependent FV strong interaction as $C^L(E^*)$,
this correlation function is
\begin{eqnarray}
&& S^\dagger \left[\ 
J_0^L(E^*)
\ +\ 
C^L(E^*) \left( J_0^L(E^*) \right)^2
\ +\ 
C^L(E^*)^2 \left( J_0^L(E^*) \right)^3
\ +\ ...
\ \right] S
\nonumber\\
& = & 
S^\dagger { J_0^L(E^*)
\over 
1 - C^L(E^*)  J_0^L(E^*)
}
S
\ =\ 
S^\dagger { 1
\over 
1/J_0^L(E^*) - C^L(E^*)
}S
\ \ \ .
\end{eqnarray}
Therefore, the FV QC that determines the $A_1^+$ energy eigenvalues is simply
\begin{eqnarray}
{1\over C^L(E^*)} & = &  J_0^L(E^*) 
\ .
\label{eq:bubbleQC}
\end{eqnarray}
In the infinite volume limit,  the Feynman diagrams represented in Fig.~\ref{fig:BubblesAndCoulomb} (given to all orders in Eq.~(\ref{Cr2b})), 
after performing the energy integrations, give
\begin{eqnarray}
J^\infty_0(E^*) & =&  -\,M\!\int\!{d^3 q\over (2\pi)^3}{1 \over q^2 - p^2}  \nonumber\\
&&+\ 4\pi\alpha M^2\!\int\!{d^3 q\over (2\pi)^3}\!\int\!{d^3 k\over (2\pi)^3} 
{1\over  q^2 - p^2}{1\over k^2 - p^2}{1\over  |{\bf q}-{\bf k}|^2} \ + \ \ldots,
\label{J0PTcont}
\end{eqnarray}
which in a FV, and using a momentum cut off, takes the form
\begin{eqnarray}
J_0^L(E^*)  & =  & 
-\,{M\over 4\pi^2 L}
\sum_{\bf n}^{\Lambda_n} { 1 \over |{\bf n}|^2-\tilde p^2} \nonumber\\
&& +\ 
{\alpha M^2\over 16\pi^5}\;\sum^{\Lambda_n}_{\bf n}\!\sum^{\infty}_{{\bf m}\neq{\bf n}}
{1 \over |{\bf n}|^2-\tilde p^2}
{1\over |{\bf m}|^2-\tilde p^2}
{1\over |{\bf n}-{\bf m}|^2} 
\ + \ \ldots,
\label{eq:J0FV}
\end{eqnarray}
where ${\tilde p}=Lp/2\pi$ and $\Lambda_n=L\Lambda/2\pi$ with
$\Lambda$ a momentum cutoff, and the ellipses signify omitted 
${\cal O}(\alpha^2)$ effects.  Note that the zero mode has been removed
from the photon propagator by the condition ${\bf m}\neq{\bf n}$.  As
the FV does not alter the ultraviolet (UV) behavior of the sums from
that of the infinite-volume integrals, the renormalization of
divergences in FV is the same as in infinite volume.  In
Eq.~(\ref{eq:J0FV}), the infinite volume hadron mass, $M$, has been
used, rather than $M^L$.  As the present analysis assumes $ML\gg 1$,
and the mass does not explicitly appear in the leading QC, this
difference represents a higher order effect.  In order to regulate the
divergent sums for numerical evaluation, while maintaining the
mass-independent renormalization scheme, Eq.~(\ref{eq:bubbleQC})
becomes~\cite{Beane:2003da}
\begin{eqnarray}
\frac{1}{C^L(E^*)}\ -\ {\rm Re}(J_0^{\infty\lbrace DR\rbrace}(E^*))  \ =\  J^L_0(E^*) \ -\ {\rm Re}(J_0^{\infty\lbrace \Lambda\rbrace}(E^*)) \ ,
\label{Cr15b}
\end{eqnarray}
where the $\textstyle{\{\}}$ superscript indicates regularization scheme, and it is straightforward to show that
\begin{eqnarray}
{\rm Re}(J_0^{\infty\lbrace \Lambda\rbrace}(E^*)) \ =\ -\frac{M\Lambda}{2\pi^2}-\frac{\alpha M^2}{4\pi}\ln\left(\frac{2 p}{\Lambda}\right)\  + \ \ldots,
\end{eqnarray}
and
\begin{eqnarray}
{\rm Re}(J_0^{\infty\lbrace DR\rbrace}(E^*)) 
\ =\ - \frac{\alpha M^2}{4\pi} \Bigg\lbrack \frac{1}{2}\gamma_E -\frac{1}{\epsilon}-1 +\ln\left(\frac{2 p}{\mu}\right)-\ln\sqrt{\pi} \Bigg\rbrack \ + \ \ldots,
\end{eqnarray}
which matches the perturbative expansion of the all-orders propagator given in Sec.~\ref{sec:rev}. 
The remaining task is to relate the FV interactions, $C^L(E^*)$, to their infinite volume counterparts, $C(E^*)$, which define the 
scattering matrix in Eq.~(\ref{eq:TSCere}),
and hence to the scattering parameters.

In general, the FV interactions, $C^L(E^*)$, result from a summation
of all bubble diagrams of the type shown in
Fig.~\ref{fig:BubblesAndRadiationA}, in which the photon is exchanged
between bubbles, or between an interaction and a bubble, or between
interactions, or produces a loop from the same interaction.
Consider a generic diagram with a photon across bubbles, as in
Fig.~\ref{fig:BubblesAndRadiationA}.  As a single bubble with CoM
kinetic energy $T^*$ scales as $\sim M\sqrt{M T^*}$, the contribution from the
photon pole is $\sim\sqrt{|{\bf p}|} \sim \sqrt{M/L}$.  Therefore,
diagrams with photons across $n$-bubbles are suppressed by $\sim
(\sqrt{M/L})^n$~\footnote{ The diagrams in
  Fig.~\ref{fig:BubblesAndRadiationA} are analogous to those involving
  radiation pions in NNEFT~\cite{Kaplan:1998tg,Kaplan:1998we}, which
  were analyzed in detail in Ref.~\cite{Mehen:1999hz}.  }.  To
determine the parametric contributions from these diagrams, it is
sufficient to evaluate the diagram without bubbles between the
insertions of the photon vertices, i.e. the photon across a single
$C(E^*)$ vertex.  Analogous arguments apply to the diagrams with
photons emerging from the strong interaction (by gauge invariance) and
connecting to bubbles, as in Fig.~\ref{fig:BubblesAndRadiationA}(c),
or other interactions.  It follows that $C^L(E^*)$ differs from
$C(E^*)$ by $\delta C^{(FV)}(E^*) = C^L(E^*) - C(E^*)$,
\begin{eqnarray}
\delta C^{(FV)}(E^*) & = & 
-\alpha \left(\ {2 a_C \over \pi M} \ \alpha_{3/2} \ +\ {4 a_C^2 r_0\over L}\ {\cal I}\ \ +\ ....\ \right)
\ ,
\end{eqnarray}
where $\alpha_{3/2}$ is a numerical constant given in the Appendix.
As these contributions depend explicitly on the scattering parameters
and do not constitute a simple multiplicative renormalization of
$p\cot\delta$, they explicitly preclude a direct extraction of
T-matrix elements.  This should come as no surprise, as the QED
interactions of systems containing two or more hadrons (or
interactions of such systems with other types of probes) are not
described by the two-body scattering parameters alone.  For instance,
in the case of two nucleons, there will be contributions from the
gauge-invariant two-body operators that contribute to the deuteron
quadrupole moment, and from the operators contributing to the electric
and magnetic polarizabilities.

It follows from Eq.~(\ref{Cr7b}) that, at ${\cal O}(\alpha)$, the
truncated $A_1^+$ FV QC for fields subject to spatial PBCs is
\begin{eqnarray}
 - {1\over a_C^\prime} \ +\ \frac{1}{2}r_0^\prime p^2 \ +\ ...
 & = & \frac{1}{\pi L}{\cal S}^C\left(\tilde p\right) \ +\ {\alpha M}\left[\ln\left({4\pi\over\alpha M {\rm  L}}\right) - \gamma_E \right]  
 \ +\ ...
\ ,
\label{barf22}
\end{eqnarray}
where the single sum over integer triplets, which determines the effects of the  strong interactions
in the absence of QED interactions, is modified to
\beq
{\cal S}^C\left(\,{x}\, \right)\ \ \equiv\  {\cal S}\left(\,{x}\, \right) \ -\ 
\frac{\alpha M L}{4\pi^3}\;{\cal S}_2\left(\,{x}\, \right) 
\ +\ 
{\alpha \ M a_C^2 r_0 \over \pi^2 L^2} {\cal I} 
\left[ {\cal S}\left(\,{x}\, \right)  \right]^2
\ +\ ...
\ ,
\label{Cr23b}
\eeq
with
\begin{eqnarray}
{\cal S}\left(\,{x}\, \right)\ & \equiv &   \sum^{\Lambda_n}_{\bf n}{1\over |{\bf n}|^2-x^2}\ -\  4 \pi\Lambda_n  \ ; \nn\\
{\cal S}_2\left(\,{x}\, \right)\ & \equiv &  
\sum^{\Lambda_n}_{\bf n}\!\sum^{\infty}_{{\bf m}\neq{\bf n}}{1\over |{\bf n}|^2-x^2}{1\over |{\bf m}|^2-x^2}{1\over |{\bf n}-{\bf m}|^2}\ -\  4 \pi^4 \ln \Lambda_n  
\ .
\end{eqnarray}
The scattering parameters in Eq.~(\ref{Cr23b}) are unprimed and 
the ellipses denote terms that are higher order in the $\alpha$, $1/L$ and $1/M$ expansions and in the ERE.
Eq.~(\ref{barf22}) is the main result of this paper.

The numerical evaluation of the function ${\cal S}(x)$ through exponential acceleration techniques
is well known~\cite{Luscher:1986pf,Luscher:1990ux}, and 
it is convenient to express the ${\cal O}(\alpha)$ regulated double sum as
\begin{eqnarray}
{\cal S}_2\left({x} \right) 
& = & 
{\cal R}
 - 
{2\over x^2}
\sum_{{\bf n}\ne {\bf 0}} {1\over |{\bf n}|^2 }{1\over |{\bf n}|^2 -x^2 }
\nonumber\\
& + & 
\sum_{{\bf n}\neq 0}\sum_{{\bf m}\neq 0,{\bf n}}
\left[
 {1 \over |{\bf n}|^2-x^2}
{1\over |{\bf m}|^2- x^2}
\! - \!
 {1 \over |{\bf n}|^2}
{1\over |{\bf m}|^2}
\right]\!\!
{1\over |{\bf n}-{\bf m}|^2}
\ \ \ ,
\label{eq:coulsum}
\end{eqnarray}
where
\begin{eqnarray}
{\cal R}\ & \equiv & \ \sum^{\Lambda_n}_{{\bf n}\neq 0}\ \sum^{\infty}_{{\bf m}\neq 0,{\bf n}}
\ {1\over |{\bf n}|^2 |{\bf m}|^2}{1\over |{\bf n}-{\bf m}|^2}\ -\ 4\pi^4\ln\Lambda_n 
\ =\ 
-178.42(01)
\ .
\label{fundsumpre}
\end{eqnarray}
The evaluation of this geometric constant, ${\cal R}$, is presented in the Appendix.

The QC given in Eq.~(\ref{barf22}) determines the FV energy eigenvalues of two like-charged hadrons.
The analogous QC for oppositely-charged hadrons can be determined from Eq.~(\ref{barf22}) 
by the substitution $\alpha\rightarrow -\alpha$ except in the argument of the logarithm where $\alpha\rightarrow +\alpha$.

\subsection{Renormalization Group Evolution}
\noindent 
It is tempting to combine the terms in brackets on the left and right
hand sides of Eq.~(\ref{barf22}), however, it is only this
particular decomposition that allows an ERE of the left hand
side~\cite{Bethe:1949yr}.  Indeed, the appearance of the logarithm on
the right-hand side is essential to the physical interpretation of the
QC, and can be understood with the aid of the renormalization group
(RG).  In the $\overline{MS}_{FV}$ scheme, a running scattering length
can be defined,
\beq
{1\over a(\mu)}\  \equiv \  {4\pi\over M C(0;\mu)} \ =\ {1\over a_C}  +  \alpha M \left[\ln\left({2\mu\over\alpha M}\right) - \gamma_E \right] \ ,
\label{Cr11b}
\eeq
which, by construction, satisfies
\beq
{1\over a(\mu)}\  = \ {1\over a(\nu)}\   +  \alpha M \ln\left({\mu\over\nu}\right) \ .
\label{Cr12}
\eeq
This (scheme-dependent) running scattering length can be interpreted as the scattering length
with the leading QED effects from distance scales $>1/\mu$ removed~\cite{Kong:1999sf}. 

With this running scattering length in mind, it is convenient to give
alternate forms of the QC, Eq.~(\ref{barf22}).  For instance, the QC
can be expressed in terms of the $\overline{MS}_{FV}$ scattering
length with the leading QED effects from length scales outside of the
spatial volume removed.  To this end, a renormalization-scale
dependent function, $\bar\delta(p,\mu)$, can be defined such that
\beq
p\cot\bar\delta(p,\mu)\ \equiv \ - {1\over a(\mu)} \ +\ \frac{1}{2}r_0 p^2 \ +\ \ldots 
\label{pcotdelbardef}
\ \ \ ,
\eeq
leading to
\beq
p\cot\bar\delta^\prime(p,2\pi/L)\ \equiv \ - {1\over a^\prime(2\pi/L)} \ +\ \frac{1}{2}r_0^\prime p^2 \ +\ \ldots 
\ =\ \frac{1}{\pi L}{\cal S}^C\left(\tilde p\right)  \ ,
\label{barfagain}
\eeq 
where the primes denote the modified kinematics.  Despite the presence
of the scheme-dependent scattering length, this form of the QC is the
most physical, as it is written only in terms of quantities which have
support within the boundaries of the FV.  The price that is paid for
expressing the QC directly in terms of the physical scattering length
is the presence of the extra term (in brackets) on the right side of
Eq.~(\ref{barf22}), which removes contributions to the scattering
length from length scales outside of the FV~\footnote{Similar
  considerations apply to the analogous QC (without EM) in two spatial
  dimensions~\protect\cite{Beane:2010ny}.}. Working with the running
scattering length, this logarithm can be absorbed, and the QC can be
expressed in terms of quantities that have support only within the FV,
i.e. $a(2\pi/L)$.  When working directly with physical
quantities, the infrared scale $\bar\mu=\alpha M e^{\gamma_E}/2$
can be chosen, which implies $a(\bar\mu)=a_C$ and the function
$p\cot\bar\delta(p)\ \equiv \ p\cot\bar\delta(p,\bar\mu)$ can be used
in the QC,
\beq
p\cot\bar\delta^\prime(p)\ = \ - {1\over a_C^\prime} \ +\ \frac{1}{2}r_0^\prime p^2 \ +\ \ldots \ =\ \frac{1}{\pi L}{\cal S}^C\left(\tilde p\right)  
\ +\ {\alpha M}\left[\ln\left({4\pi\over\alpha M L}\right) - \gamma_E \right]  \ .
\label{marform}
\eeq

\subsection{Approximate Energy Eigenvalues}
\noindent
Ideally, the QC in Eq.~(\ref{barf22}) is solved numerically to
determine the FV energy eigenvalues.  However, the smallness of
$\alpha M L$ in present day calculations, and those of the foreseeable
future, implies that the QED FV shifts in the two-hadron energy
eigenvalues are small, and the numerics will not be particularly
enlightening.  However, considering the ${\cal O}(\alpha)$
perturbative corrections to the eigenvalues is informative.  It is
worth emphasizing the somewhat peculiar nature of the expansions in
the approximate formulas that follow, which suggest a somewhat narrow
range of validity.  While the expansions are formally perturbative in
$1/L$ times the length scale which characterizes the strength of the
interaction, and are also nonrelativistic, it is further assumed that
$M L \ll 1/\alpha$ so that the QED interactions can be treated
perturbatively.  

\subsubsection{The Ground State }
\noindent 
In a perturbative expansion around the non-interacting ground state,
with energy $E=2 M^L$, there is no contribution from the QED
interactions at ${\cal O}(\alpha)$ in the absence of strong
interactions.  This is due to the absence of the photon zero mode,
with the uniform background charge density in the unperturbed state
exactly canceling the particle charge density~\footnote{ Note that the
  ground-state energy of boosted systems will have pure Coulomb
  corrections as the charge density is no longer uniformly zero.  }.
Using standard methods, it is straightforward to find the ground-state
energy shift for scattering parameters that are small compared to $L$,
\begin{eqnarray}
  \label{eq:apprEzeroa}
\hskip-.9cm 
\Delta E_0^C
\!\! &=&
\Delta E_0
 + \Delta E_0^{(\alpha)}
 \nonumber\\
& = & \!\!\frac{4\pi\, a^\prime }{M\,L^3}\Bigg\{\!1
-\left(\frac{a^\prime}{\pi\,L}\right){\cal I}
+\left(\frac{a^\prime}{\pi\,L}\right)^2\left[{\cal I}^2-{\cal J}\right]
+
\ldots
\Bigg\}  \nn \\
&&
-\ \frac{2\alpha\, a^\prime}{L^2 \pi^2}\Bigg\{\! {\cal J}
+\left(\frac{a^\prime}{\pi\,L}\right)\left[{\cal K}- {\cal I}{\cal J}-{\cal R}/2 \right]
\nonumber\\
&&\qquad\qquad
+\left(\frac{a^\prime}{\pi\,L}\right)^2\left[{\cal R}{\cal I}+ {\cal I}^2{\cal J}-2{\cal J}^2-2{\cal I}{\cal K}+{\cal L}-{\cal R}_{24}\right]\ 
\nonumber\\
&&\qquad\qquad
+{2 a^\prime r_0^\prime \pi^2\over L^2} {\cal I}
 + \dots\Bigg\} \ ,
\end{eqnarray}
where $a^\prime\equiv a^\prime(2\pi/L)$ is the $\overline{MS}_{FV}$
scattering length and the geometric constants, ${\cal I}$, ${\cal J}$,
${\cal K}$, ${\cal L}$, ${\cal R}$ and ${\cal R}_{24}$ are defined and
evaluated in the Appendix.  The first term in braces is the well-known
energy shift due to QCD interactions, while the second term
is the shift due to the combined QCD and QED interactions.  This can
also be expressed in terms of the kinematically-shifted  scattering parameters,
\begin{eqnarray}
  \label{eq:apprEzeroac}
\hskip-.9cm 
\Delta E_0^C
&=&
\!\!\frac{4\pi\, a_C^\prime}{M\,L^3}\Bigg\{\!1
-\left(\frac{a_C^\prime}{\pi\,L}\right){\cal I}
+\left(\frac{a_C^\prime}{\pi\,L}\right)^2\left[{\cal I}^2-{\cal J}\right]
+
\ldots
\Bigg\}  \nn \\
&&-
\ \frac{2\alpha\, a_C^\prime}{L^2 \pi^2}\Bigg\{\! {\cal J}
+\left(\frac{a_C^\prime}{\pi\,L}\right)\left[{\cal K}- {\cal I}{\cal J}-{\tilde{\cal R}}/2 \right]
\nonumber\\
&&\qquad\qquad
+\left(\frac{a_C^\prime}{\pi\,L}\right)^2\left[{\tilde{\cal R}}{\cal I}+ {\cal I}^2{\cal J}-2{\cal J}^2-2{\cal I}{\cal K}+{\cal L}-{\cal R}_{24}\right] \ 
\nonumber\\
&&\qquad\qquad
+ {2 a_C^\prime r_0^\prime \pi^2\over  L^2} {\cal I}
+ \dots\Bigg
\} 
\ ,
\end{eqnarray}
where  
\begin{eqnarray}
{\tilde{\cal R}}\ \equiv\ {\cal R}\ -\ 4\pi^4 \left[\ln\left({4\pi\over\alpha M L}\right) - \gamma_E \right] \ .
\label{tildeR}
\end{eqnarray}
The ellipsis denote terms that are higher order in $1/M$, $1/L$ and $\alpha$.
In terms of the physical scattering parameters, the energy shift of the ground state is
\begin{eqnarray}
  \label{eq:apprEzeroacup}
\hskip-.9cm 
\Delta E_0^C
&=&
\!\!\frac{4\pi\, a_C}{M\,L^3}\Bigg\{\!1
-\left(\frac{a_C}{\pi\,L}\right){\cal I}
+\left(\frac{a_C}{\pi\,L}\right)^2\left[{\cal I}^2-{\cal J}\right]
+
\ldots
\Bigg\}  \nn \\
&&-
\ \frac{2\alpha\, a_C}{L^2 \pi^2}\Bigg\{\! {\cal J}
+\left(\frac{a_C}{\pi\,L}\right)\left[{\cal K}- {\cal I}{\cal J}-{\tilde{\cal R}}/2 \right]
\nonumber\\
&&\qquad\qquad
+\left(\frac{a_C}{\pi\,L}\right)^2\left[{\tilde{\cal R}}{\cal I}+ {\cal I}^2{\cal J}-2{\cal J}^2-2{\cal I}{\cal K}+{\cal L}-{\cal R}_{24}\right] \ 
\nonumber\\
&&\qquad\qquad
+ { a_C r_0 \pi^2\over  L^2} {\cal I}
+ \dots\Bigg
\} 
\ ,
\end{eqnarray}
The only difference 
between Eq.~(\ref{eq:apprEzeroac}) and Eq.~(\ref{eq:apprEzeroacup})
is the coefficient of the last term, as other differences are higher order in the expansion.

\subsubsection{The First Excited State }
\noindent
In contrast to the ground state, the energy shift of the first excited
state in the FV receives a contribution from the
exchange of a single Coulomb photon as the uniform background charge density
does not cancel against the $|{\bf n}|=1$ unperturbed two-hadron
charge density.  Following
L\"uscher~\cite{Luscher:1986pf,Luscher:1990ux} and
expanding~\footnote{Note that this expansion requires special care due
  to the singular, purely Coulombic, contribution.} the energy shift
in terms of $\tan\overline{\delta}^\prime$ evaluated at the
unperturbed energy, the energy shift of the first excited state is
\begin{eqnarray}
  \label{eq:apprEonea}
\hskip-.9cm 
\Delta E_1^C
\!\! &=&
\Delta E_1
 + \Delta E_1^{(\alpha)}
 \nonumber\\
& = & 
{4\pi^2\over M L^2}
\ -\ {12 \tan\overline{\delta}^\prime\over M L^2}
\left( 1 + c_1^\prime \tan\overline{\delta}^\prime + c_2^\prime \tan^2 \overline{\delta}^\prime + ... \right)
\nonumber\\
&& \ +\ 
{9 \alpha\over 4 \pi {\rm  L} } \left( 1 + c_{1\alpha}^\prime  \tan\overline{\delta}^\prime 
+\left(c_{2\alpha}^\prime 
+ {8\over 3}  \log\left(\alpha M L\right) \right) \tan^2\overline{\delta} ^\prime
+ ... \right)
\ \ \ ,
\end{eqnarray}
where $c_1^\prime = -0.061365$, $c_2^\prime = -0.35415$ and 
$c_{1\alpha}^\prime = 3.83582$, 
$c_{2\alpha}^\prime =  -7.12197$,
The strong coefficients, $c_{1,2}^\prime$ were first computed by L\"uscher~\cite{Luscher:1986pf,Luscher:1990ux},
and we do not repeat their determination here.
The leading QED contribution arises from
the exchange of a single Coulomb photon between $|{\bf n}|=1$ two-hadron states, and is simply given by
\begin{eqnarray}
&& {\alpha\over 6\pi L} 
 \sum_{|{\bf m}|,|{\bf n}|=1\atop {\bf n}\ne {\bf m}} {1\over |{\bf n}-{\bf m}|^2}
 \ = \ {9\alpha\over 4\pi L}
\ \ \ ,
\end{eqnarray}
while the remaining QED contributions are of the form
\begin{eqnarray}
c_{1\alpha}^\prime &=& -\frac{4}{9\pi^2}\left( 6-{\cal X}_2 \right) \ \ ; \nn\\
c_{2\alpha}^\prime &=& -\frac{2}{3\pi^4}\Big\lbrack 
\frac{1}{3}\left( 6-{\cal X}_2 \right) {\cal I}^{(1)}+ \frac{1}{6}{\cal X}_1 {\cal J}^{(1)}-{\cal R} +12 \nn\\
&&\qquad\qquad\qquad\qquad+2\left({\cal X}_3+{\cal X}_4+{\cal X}_5\right)+ {\cal X}_1-{\cal X}_6 -4\pi^4\left(\gamma_E -\log 4\pi \right) \Big\rbrack 
\ \ \ ,
\end{eqnarray}
where the geometric constants, ${\cal I}^{(1)}$, ${\cal J}^{(1)}$, and
${\cal X}_1$-${\cal X}_6$ are defined and evaluated in the Appendix.
Terms higher order in $1/L$, such as the leading contribution from
$r_0$ at $1/L^2$ , e.g. $+ {9 \alpha\over 4 \pi L} {3 r_0\over \pi L}
\tan^2\overline{\delta} $, are not shown.  Further, at this order,
$\tan\overline{\delta}^\prime$ can be replaced with $\tan\overline{\delta}$
without modifying the form of Eq.~(\ref{eq:apprEonea}).  To give some
perspective, in a ${\rm L}=10~{\rm fm}$ volume, the leading ${\cal
  O}(\alpha)$ energy shift from the exchange of a single Coulomb
photon is $\sim 100~{\rm keV}$.

\subsubsection{The (Possible) Bound State }
\noindent
In nature, there are no bound doubly-charged two hadron systems, however
such systems do exist at unphysical pion masses, as determined with
Lattice QCD calculations~\cite{Beane:2011iw,Yamazaki:2012hi,Beane:2012vq,Beane:2013br}.  The bound-state energy
in the FV is determined from the large-$x$ limit of ${\cal S}^C(x)$ in
Eq.~(\ref{barf22}) and Eq.~(\ref{Cr23b}), and, in particular, the sums
contributing to ${\cal S}^C(x)$ are
\begin{eqnarray}
 \sum_{\bf n}^{\Lambda_n} { 1 \over |{\bf n}|^2+\tilde \kappa^2}
& \rightarrow &
4\pi\Lambda_n \ -\ 2 \pi^2 \tilde \kappa 
\ \ ;
\label{eq:bssumsa}\\
 \sum_{\bf n}^{\Lambda_n} 
 \sum_{\bf m\ne n} 
{1\over |{\bf m}|^2+\tilde \kappa^2}
{1\over |{\bf n}|^2+\tilde \kappa^2}
{1\over |{\bf n}-{\bf m}|^2}
& \rightarrow &
4\pi^4\ \left( \log\Lambda_n\ -\ \log\left(2 \tilde \kappa\right)\ \right)
\ +\ 
{\pi^2\over\tilde\kappa} {\cal I}
\ \ \ ,
\label{eq:bssumsb}
\end{eqnarray}
in the large volume limit, $\tilde\kappa\rightarrow\infty$, 
where only the leading power-law corrections are shown~\footnote{Eq.~(\ref{eq:bssumsb}) is
obtained by first shifting ${\bf m}\rightarrow{\bf n}+{\bf p}$, performing the sum over ${\bf n}$ using
the Poisson summation formula, and then dividing the sum over ${\bf p}$ into two regions. The first
region generates the power law correction, and the second region is again evaluated using Poisson summation
to give the logarithmic contributions.}.  At the order
to which we are working, these limits lead to a FV QC for the bound
state of
\begin{eqnarray}
-{1\over a_C} - {1\over 2} r_0 \kappa^2
& = & - \kappa - \alpha M \left( \gamma_E + \log\left({\alpha M\over 4\kappa}\right)\ \right)
\ -\ {\alpha M \over 2\pi \kappa L} (1-\kappa r_0) {\cal I}
\ \ \ ,
\end{eqnarray}
which determines the leading Coulomb corrections to the bound-state binding energy.
Performing a perturbative expansion of $\kappa = \kappa_0 + \kappa_1 + ...$
leads to a binding energy of
\begin{eqnarray}
B^C  & = & 
{\kappa_0^{ 2}\over M}
\ -\ 
{2\alpha \kappa_0 \over 1-\kappa_0 r_0 }\ \left[\ 
\gamma_E\ +\ \log\left({ \alpha M\over 4 \kappa_0}\right)
\ \right]
\ -\ 
{\alpha \over\pi L} {\cal I}
\ +\ ...
\ \ \ ,
\end{eqnarray}
where $\kappa_0$ is the binding momentum resulting from the strong
interactions alone.  The leading QED contribution to the
infinite-volume binding energy is consistent with a direct
perturbative calculation in the ER theory.  Further, the leading FV
correction to the binding is given~\footnote{The relation between the
scattering parameters and the binding momentum has been used, with
terms higher order in the scattering parameters neglected.}, which
vanishes as $1/L$, as expected.

In the limit in which the bound state is compact compared to the
lattice volume, the leading corrections to its total mass should be
that of a charge-2 system, as calculated in
Ref.~\cite{Davoudi:2014qua}.  There are two contributions to the mass
shift of the bound state, one from the shifts of the individual
constituent hadrons, and one from the shift in the binding energy.  We
find that in the deep-binding limit, the total mass of the bound state
is shifted by
\begin{eqnarray}
\delta M_{BS}^{(FV)} & = & 2 \delta M^{(FV)} - \delta B^C
\ =\ 
2 \left( {\alpha\over 2\pi L} {\cal I}\right) + {\alpha \over\pi L} {\cal I}
\ +\ ...
\ =\ {2 \alpha \over\pi L} {\cal I}
\ +\ ...
\ \ \ ,
\end{eqnarray}
consistent with expectations~\cite{Hayakawa:2008an,Davoudi:2014qua}.

\section{Summary and Discussion}
\noindent 
Lattice QCD has reached the point where QED is being included in
calculations of some of the simplest hadronic properties, such as the
masses of the lowest lying hadrons.  Naively, the inclusion of QED
should be problematic for calculations in a finite volume due to its
long range nature.  However, by simply omitting the zero-modes of the
photon field, which lead to the violation of both Gauss's and Ampere's
laws, Lattice QCD+QED calculations can be performed in meaningful ways
to reliably extract important quantities without corrupting the
infinite-volume limit.  Recently, the relation between the single
hadron masses calculated in a finite volume and their infinite-volume
values has been established~\cite{Hayakawa:2008an,Davoudi:2014qua}.
Given the nonperturbative nature of the Coulomb interaction in
low-energy scattering, extending this work to relate two-hadron energy
eigenvalues to their corresponding S-matrix elements had the potential
to be quite involved.  In this work, we have shown that there is a
large range of volumes, satisfying $ML \ll 1/\alpha$, for which the
non-relativistic relation between the finite-volume energy of two
hadrons in the $A_1^+$ representation of the cubic group and their
s-wave phase shift receives calculable perturbative QED corrections.
Our results will straightforwardly generalize to the relations between
the energies of two hadrons in other representations of the cubic
group and the phase shifts and mixing parameters in all relevant
scattering channels.

The confining nature of QCD simplifies the evaluation of hadronic
correlation functions using Lattice QCD, as it dictates that the
interactions among hadrons are contained within a volume set by the
longest correlation length, which is the pion Compton wavelength.  As
long as the size of the spatial lattice is significantly larger than
the inverse pion mass, there is a hierarchy of length scales and
finite-volume artifacts can be removed, as in the case of
single-particle properties, or exploited, as in the calculation of
two-particle interactions.  The presence of an infinite-range force
destroys this hierarchy.  With no zero modes and a gap in the spectrum
of the momentum operator, there is a region of parameter space for the
calculation of the energy of two non-relativistic hadrons of mass
$M$. In particular, if the lattice volume satisfies $ML \ll 1/\alpha$,
Coulomb ladders are perturbative, and their contribution to the
two-particle energy, along with other contributions that are absent in
infinite volume, can be computed perturbatively in
$\alpha$. Furthermore, in the absence of zero modes, the gap in the
spectrum sets the scale of the contribution due to inelastic
processes.  It is essential that such a gap exist in order to derive
the quantization conditions that relate the energies computed in LQCD
and relevant S-matrix elements - those dictating the two-hadron
scattering amplitude, and those which determine electromagnetic
processes.  In the absence of QED, the low-energy EFT, which is valid
up to the start of the QCD t-channel cut, gives a QC in a form that is
valid up to the QCD inelastic threshold when expressed in terms of
$p\cot\delta$ (see Fig.~\ref{fig:panalyticstrong}).  However, it is
important to stress that in the presence of QED, the expressions we
have derived are valid up to the QED inelastic threshold when this
lies below the QCD t-channel cut, or otherwise up to the QCD t-channel
cut.

\section*{\hspace{-1.1cm} Acknowledgments}
\noindent 
We would like to thank Zohreh Davoudi and David B.~Kaplan for
useful discussions.  SRB was supported in part by NSF continuing
grant PHY1206498 and MJS was supported in part 
by DOE grant No.~DE-FG02-00ER41132.

\renewcommand{\theequation}{A-\arabic{equation}}
\setcounter{equation}{0}  

\section*{\hspace{-1.1cm} APPENDIX:   Integer Sums}  

\subsection*{\hspace{-1.1cm}Single Sums}  
\noindent 
The single sums over integer triplets that are required  for the modified kinematics in a finite volume and 
for the approximate two-hadron energy eigenvalues are:
\begin{eqnarray}
  {\cal I}&=&\sum_{{\bf n}\neq 0}^{\Lambda_n} \frac{1}{|{\bf n}|^2} -4\pi\Lambda_n = -8.9136
\qquad\qquad  , \qquad\qquad  
  {\cal J}\ =\ \sum_{{\bf n}\neq 0} \frac{1}{|{\bf n}|^4}  = 16.5323  \ ; 
  \nn \\
  {\cal K}&=&\sum_{{\bf n}\neq 0} \frac{1}{|{\bf n}|^6}  = 8.4019
\qquad\qquad\qquad  , \qquad\qquad\qquad
  {\cal L}\ =\ \sum_{{\bf n}\neq 0} \frac{1}{|{\bf n}|^8}  =  6.9458 \ ; 
  \nn\\
{\cal I}^{(1)} &=&  \sum_{ |{\bf n}|\ne 1}  {1\over |{\bf n}|^2-1} 
\ =\  {-1.2113 } 
\qquad\qquad  , \qquad\qquad  
{\cal J}^{(1)} \ =\   \sum_{ |{\bf n}|\ne 1}  {1\over (|{\bf n}|^2-1)^2}  \ =\  {23.2430} \  ;
  \nn\\
  {\cal X}_1 & =& \sum_{|{\bf m}|,|{\bf n}|=1\atop {\bf n}\ne {\bf m}} {1\over |{\bf n}-{\bf m}|^2}
 \ =\   { {27\over 2}  }
\quad\qquad , \qquad\quad
{\cal X}_2\ =\ \sum_{ |{\bf n}|=1\atop |{\bf m}|> 1} {1\over |{\bf m}|^2-1} {1\over |{\bf n}-{\bf m}|^2}
\  = \ {91.1806} \  ;
\nn\\
{\cal X}_3 & =& \sum_{ |{\bf n}| > 1}  {1\over  |{\bf n}|^2 ( |{\bf n}|^2-1)} 
\ = \ 
{14.7022} \qquad , \qquad 
{\cal X}_4\ =\  \sum_{ |{\bf m}|=1\atop  |{\bf n}| > 1}   {1\over |{\bf n}|^2}   {1\over |{\bf n}-{\bf m}|^2}
\ =\ 
{65.3498} \  ;
\nn\\
{\cal X}_5 &=& \sum_{ |{\bf n}|=1\atop |{\bf m}|> 1} {1\over (|{\bf m}|^2-1)^2} {1\over |{\bf n}-{\bf m}|^2}
\ =\ 
{46.5687}
\ \ \ .
\end{eqnarray}
%

\subsection*{\hspace{-1.2cm}Double Sums}  
\noindent 
Unlike the situation  in  large volumes when only strong interactions contribute, 
and explicit two-loop sums are not required,
the leading QED contributions resulting from the exchange of Coulomb photons require non-trivial two-loop sums over triplets of integers.
Consider the finite double sum:
\begin{eqnarray}
{\cal R}\ 
& \equiv &
\sum^{\Lambda_n}_{{\bf n}\neq 0}\ \sum^{\infty}_{{\bf m}\neq 0,{\bf n}}\ {1\over |{\bf n}|^2 |{\bf m}|^2}{1\over |{\bf n}-{\bf m}|^2}\ -\ 4\pi^4\ln\Lambda_n 
\nonumber\\
& = & 
\sum^{\Lambda_n}_{{\bf n}\neq 0}\ \ {1\over |{\bf n}|^2}\ {\cal R}_{sub}({\bf n})\
\ -\ 4\pi^4\ln\Lambda_n 
\ .
\label{fundsum}
\end{eqnarray}
It is regulated asymmetrically, by first evaluating the inner sum without a cut off, 
\begin{eqnarray}
{\cal R}_{sub}({\bf n})\ \equiv\ \sum^{\infty}_{{\bf m}\neq 0,{\bf n}}\ {1\over |{\bf m}|^2}{1\over |{\bf n}-{\bf m}|^2} 
\ ,
\label{fundsumsub}
\end{eqnarray}
and then straightforwardly evaluated using the methods described in Ref.~\cite{Tan:2007bg}.
It is found to be
\begin{eqnarray}
{\cal R}_{sub}({\bf n})& =& -2\eta \left(1-e^{-\eta |{\bf n}|^2}\right)\frac{1}{|{\bf n}|^2} 
\ +\ \sum^{\infty}_{{\bf m}\neq 0,{\bf n}}\ \left(e^{-\eta D_{{\bf n}{\bf m}}}+e^{-\eta |{\bf m}|^2}
-e^{-\eta (D_{{\bf n}{\bf m}}+|{\bf m}|^2)}\right)\frac{1}{|{\bf m}|^2 D_{{\bf n}{\bf m}}} 
\nn \\
&+& \int d^3 {\bf m} \left(1-e^{-\eta D_{{\bf n}{\bf m}}}\right)\left(1-e^{-\eta m^2}\right)\frac{1}{|{\bf m}|^2D_{{\bf n}{\bf m}}} \ ,
\label{fundsumsubTan}
\end{eqnarray}
where $D_{{\bf n}{\bf m}}\equiv |{\bf n}-{\bf m}|^2$, and $\eta$ is a
small number introduced to provide a clean way to separate sums into
UV and IR contributions where the UV sums can be replaced by
integrals. The $\eta$ used here should not be confused with the
kinematic variable used in the main body of the paper.  The
$\eta$-independent piece (in the integral) is readily evaluated, 
giving
\begin{eqnarray}
{\cal R}_{sub}({\bf n})
& =& 
\frac{\pi^3}{|{\bf n}|} 
-2\eta \left(1-e^{-\eta |{\bf n}|^2}\right)\frac{1}{|{\bf n}|^2} 
\nonumber\\
&& +   
\sum^{\infty}_{{\bf m}\neq 0,{\bf n}}\ \left(e^{-\eta D_{{\bf n}{\bf m}}}+e^{-\eta |{\bf m}|^2}
\ -\ 
e^{-\eta (D_{{\bf n}{\bf m}}+|{\bf m}|^2)}\right)\frac{1}{|{\bf m}|^2 D_{{\bf n}{\bf m}}} 
 \nn \\
&& - 
2\pi \int_0^\infty dm \int_{-1}^1 dc \left(e^{-\eta D_{nmc}}+e^{-\eta  m^2}- e^{-\eta (D_{nmc}+  m^2)}\right)\frac{1}{D_{nmc}}  \ .
\label{fundsumsubTan2}
\end{eqnarray}
where $D_{nmc}\equiv |{\bf n}|^2-2 |{\bf n}| |{\bf  m}| c+|{\bf m}|^2$,
from which Eq.~(\ref{fundsum}) becomes
\begin{eqnarray}
{\cal R}\ =\ \pi^3\;\alpha_{3/2} \ -\ 2\eta\;{\cal J} \ +\ 2\eta\;{\cal J}^\eta \ +\  {\cal T}_1 \ -\ 2\pi\;{\cal T}_2 
\ =\ 
-178.42(01)
\ .
\label{fundsum2}
\end{eqnarray}
where~\cite{Tan:2007bg}
\begin{eqnarray}
\alpha_{3/2} \ \equiv\ \sum^{\Lambda_n}_{{\bf n}\neq 0} {1 \over |{\bf n}|^3} \ -\ 4\pi\ln\Lambda_n \ =\ 3.8219 \ , 
\label{alph4bare}
\end{eqnarray}
The $\eta$-dependent sums are
\begin{eqnarray}
&&{\cal J}^\eta \ \equiv\ \sum^{\infty}_{{\bf n}\neq 0} 
{e^{-\eta |{\bf n}|^2}\over |{\bf n}|^4}
\ \ ;
\nonumber\\
&&{\cal T}_1\ \equiv\ 
\sum^{\infty}_{{\bf n}\neq 0}\ \sum^{\infty}_{{\bf m}\neq 0,{\bf n}} 
\left(e^{-\eta D_{{\bf n}{\bf m}}}
+
e^{-\eta |{\bf m}|^2}
-
e^{-\eta (D_{{\bf n}{\bf m}}+|{\bf m}|^2)}\right)\frac{1}{|{\bf n}|^2|{\bf m}|^2D_{{\bf n}{\bf m}}} 
\ \ ;
\nonumber\\
&&{\cal T}_2 \equiv 
\sum^{\infty}_{{\bf n}\neq 0} \frac{1}{|{\bf n}|^2}\int_0^\infty dm \int_{-1}^1 dc \left(e^{-\eta D_{nmc}}+e^{-\eta m^2}- e^{-\eta (D_{nmc}+m^2)}\right)\frac{1}{D_{nmc}}  
\ ,
\label{alph4eta}
\end{eqnarray}
where are all evaluated numerically for a range of values of $\eta$ that provide stable results for each.

Evaluation of the perturbative expansion of the ground-state energy requires sums of the form
\begin{eqnarray}
{\cal R}_{st}\ \equiv\ \sum^{\infty}_{{\bf n}\neq 0}\ \sum^{\infty}_{{\bf m}\neq 0,{\bf n}}\ {1\over |{\bf n}|^s |{\bf m}|^t}{1\over |{\bf n}-{\bf m}|^2}
\ ,
\label{fundsumgen}
\end{eqnarray}
but at the order to which we have worked, only ${\cal R}_{24} = 170.97(01)$ is required.
Further,
in the perturbative expansion of the energy of the first excited states, the two-loop sum
\begin{eqnarray}
&&
{\cal X}_6\ =\ \sum_{|{\bf m}|, |{\bf n}| > 1 \atop {\bf n}\ne{\bf m}}
 \left( 
 {1\over |{\bf n}|^2-1} {1\over |{\bf m}|^2-1} - {1\over |{\bf n}|^2}{1\over |{\bf m}|^2}
 \right)
  {1\over |{\bf n}-{\bf m}|^2}
  \ =\  
{264.508}
\ \ \ ,
\end{eqnarray}
is required, and it is evaluated with techniques similar to those used previously.

\bibliography{bibi}

\end{document}